\documentclass[10pt,aps,prb,twocolumn]{revtex4-1}
\usepackage{amssymb,amsmath}
\usepackage{subfig}
\usepackage{graphicx}
\usepackage{color}
\usepackage{hyperref}
\usepackage{listings}
\usepackage{ads}
\usepackage[utf8]{inputenc}
\usepackage{CJKutf8}
\usepackage{natbib}

\hypersetup{
    bookmarks=true,         
    unicode=false,          
    pdftoolbar=true,        
    pdfmenubar=true,        
    pdffitwindow=false,     
    pdfstartview={FitH},    
    pdfauthor={Guandi Zhao},     
    colorlinks=false,       
    citecolor=red,        
}

\graphicspath{{figures/}}
\newcommand{\lcdm}{$\Lambda$CDM}

\begin{document}

\definecolor{dkgreen}{rgb}{0,0.6,0}
\definecolor{gray}{rgb}{0.5,0.5,0.5}
\definecolor{mauve}{rgb}{0.58,0,0.82}

\lstset{frame=tb,
  	language=Python,
  	aboveskip=3mm,
  	belowskip=3mm,
  	showstringspaces=false,
  	columns=flexible,
  	basicstyle={\small\ttfamily},
  	numbers=none,
  	numberstyle=\tiny\color{gray},
 	keywordstyle=\color{blue},
	commentstyle=\color{dkgreen},
  	stringstyle=\color{mauve},
  	breaklines=true,
  	breakatwhitespace=true
  	tabsize=3
}

\title{Halo spin and orientation in Interacting Dark Matter Dark Energy Cosmology}
\author{Guandi Zhao}
\email{guzhao@ethz.ch}
\affiliation{Department of Physics, ETH Zurich, 8092 Zurich, Switzerland}
\author{Jiajun Zhang}
\email{jjzhang@shao.ac.cn}
\affiliation{Shanghai Astronomical Observatory, Chinese Academy of Sciences, Shanghai 200030, China}
\author{Peng Wang}
\affiliation{Shanghai Astronomical Observatory, Chinese Academy of Sciences, Shanghai 200030, China}
\author{Ji Yao}
\affiliation{Shanghai Astronomical Observatory, Chinese Academy of Sciences, Shanghai 200030, China}
\date{\today}

\begin{abstract}

In recent years, the interaction between dark matter (DM) and dark energy (DE) has become a topic of interest in cosmology. Interacting dark matter–dark energy (IDE) models have a substantial impact on the formation of cosmological large-scale structures, which serve as the environment for DM halo evolution. This impact can be examined through the shape and spin orientation of halos in numerical simulations incorporating IDE effects. In our work, we used the N-body simulation pipeline \texttt{ME-GADGET} to simulate and study the halo spin and orientation in IDE models. We found that in models where DM transfers into DE (IDE I), the alignment of halo shapes with the surrounding tidal field is enhanced compared to \lcdm, while the alignment of halo spins with the tidal field is decreased compared to the \lcdm. Conversely, in models where DE transfers into DM (IDE II), the opposite occurs. We have provided fitted functions to describe these alignment signals. Our study provides the foundation for more accurate modeling of observations in future surveys such as China Space Station Telescope (CSST). 

 
\end{abstract}

\maketitle

\section{\label{sec:zero}Introduction}



The accelerating expansion of the current universe implies the presence of a component with negative pressure, identified as Dark Energy (DE)\cite{accelerated_expansion}. \lcdm \ model assumes constant dark energy density ($\Lambda$), and is supported by observations of both the early and late universe, such as Cosmic Microwave Background (CMB)\cite{wmap_cmb,Planck_2018,planck_collaboration_planck_2016}, Baryonic Acoustic Oscillations (BAO)\cite{SDSS_BAO,BOSS_BAO}, as well as the observation of the later universe by type Ia Supernova (SNIa)\cite{accelerated_expansion,Omega_Lambda_SnIa}.

However, \lcdm\  model is challenged by recent observations, especially the Hubble tension and the S8 tension\cite{review_Hubble_tension,freedman_2017_cosmology,2019_Reiss_H0,2021_Valetino_S8,2022_Elcio_S8}. The Hubble tension is a discrepancy between indirect measurements of early universe, such as CMB and Big Bang nucleosynthesis (BBN) and direct constraints based on local distance ladder 
\cite{BOSS_BAO,SDSS_BAO,Riess_2021,Planck_2018}, and S8 tension is the discrepancy between matter anisotropy measured by CMB and the matter clustering strength measured by weak lensing experiments \cite{2017_KiDS_S8,2020_KiDsViKing_S8,2021_KiDs1000_S8,2022_Elcio_S8,2020_DES_S8,2023_HSC_S8}.


Besides, recent results from the DESI collaboration, combining DESI (FS+BAO), CMB, and DES-SN5YR data \cite{Zhao_2017,roy2024dynamicaldarkenergylight}, provide evidence for a dynamical dark energy equation of state (EoS) ($w_0 w_a$CDM model), parameterized by $w(a) = w_0+ (1-a)w_a$ with posterior $w_0=-0.761\pm 0.065$ and $w_a=-0.96_{-0.26}^{+0.30}$. This suggests that dark energy may have a varying ratio between mass and pressure over the evolution of the universe. Though as recently pointed out by Ó Colgáin and Sheikh-Jabbari \cite{2024_Roin_DES}, the evidence for a dynamical dark energy signal may be influenced by systematic discrepancies, exploring the implications of a varying dark energy equation of state remains valuable.

Interacting dark energy–dark matter (IDE) models provide a framework to explain the evolution of the dark energy EoS and have acquired academic interest. Besides, IDE models offer a potential solution to the $H_0$-S8 tension \cite{2017_Valetino_IDE,2020_Valetino_IDE,2018_Yang_IDEH0,2024_Gao_IDE,2024_Wang_IDE}. In the future, observations conducted by telescopes such as China Space Station Telescope (CSST) may provide key evidence through weak lensing\cite{Yao2024MNRAS}. A more precise modeling of observational signals is necessary.




Cosmological simulations can provide hints for DM-DE interactions beyond linear perturbation, and relate to observations. Baldi et al. \cite{baldi_clarifying_2011} first examined the effects of IDE in nonlinear matter power spectrum using N-body simulation. In 2018, Zhang et al. \cite{zhang_fully_2018} adopted \texttt{ME-GADGET}(a modified version of the \texttt{GADGET-2} \cite{GADGET_code_1,GADGET_code_2}), and incorporated IDE effects into DM-only N-body simulation. The result demonstrated that the mass power spectrum difference across IDE models happens after $z\sim 1$. In a consecutive paper\cite{liu_dark_2022}, Liu et al. examined the halo mass function and the halo concentration-mass relation across IDE models. Zhao et al.\cite{2023MNRASZhao} did a further study of the concentration-mass relation and constrained the IDE parameters using observations. The spin and shape alignment with the large-scale environment in f(R), DGP and coupled dark energy model was studied in \cite{2017MNRASLHuillier}. 

In this work we are particularly interested in exploring the IDE influence on intrinsic alignment (IA). The intrinsic alignment (IA) signal is coherent with the cosmic shear signal, and brings systematics in weak lensing experiments such as CSST \cite{2023CSSTIA}. IA signal comes from galaxy-galaxy alignment, and galaxy alignment is correlated to the DM distribution \cite{2015IAreviewTroxel}. IA calibration modeling require a comprehensive understanding of inter-halo alignment and Large-Scale Structure (LSS) alignments \cite{2015IAreviewJoachimi}. A more detailed characterization of the IDE influence on halo-halo and halo-LSS alignment could benefit the halo model approach of IA calibration modeling.

Our work adopts the same simulation dataset and convention as Zhao et al. \cite{2023MNRASZhao}, and focuses on the examination of the relation between DM halos and their forming environments across IDE models. In this work, section \ref{method} covers the theory and simulation approach of this work; section \ref{result} lists the halo alignment relation across IDE models, section \ref{discussion} discusses the implications of halo alignment relation and possible observational probes.


\section{Methodology\label{method}}
\subsection{\label{subsec:2.1}IDE Cosmology}

IDE affects halo formation in two major aspects. Firstly, IDE models affect halo formation time; secondly, IDE models modify halo mass and thus their self and mutual gravity. The background expansion history is a modified version of the \lcdm \ spacetime with flat FLRW metric:
\begin{equation}
    ds^2 = -dt^2+a^2(t)(dx^2+dy^2+dz^2),
\end{equation}
where $a(t)$ is the scale factor. The 00 component for Einstein field equations gives the first Friedmann equation:
\begin{equation}
    \left(\frac{H}{H_0}\right)^2=\frac{8\pi G}{3} (\rho_M+\rho_E),\label{eq:Friedmann}
\end{equation}

where $H(t)=\frac{1}{a}\frac{da}{dt}$ is the Hubble parameter, $H_0$ is the current Hubble rate, $G$ is the gravitational constant, and $\rho_M,\rho_E$ are DM and DE energy density (00 component of the energy-momentum tensor).

Since the energy is conserved within the DE-DM sector:
\begin{equation}
    \nabla_\mu T^{\mu\nu}_{M}+\nabla_\mu T^{\mu\nu}_{E}=0,
\end{equation}

where $T_{M}, T_{E}$ are the energy momentum tensor of DM and DE, respectively.

In this work, we use a phenomenological model to parametrize IDE models based on the time component energy-momentum transfer $Q=Q^0$.

The energy momentum transfer between DM and DE induced by IDE models is defined as the four-vector $Q^\nu$, with the following equations:
\begin{gather}
    \nabla_{\mu}T^{\mu\nu}_{M}=Q^\nu,\label{eq:DM_T}\\
    \nabla_{\mu}T^{\mu\nu}_{E}=-Q^\nu\label{eq:DE_T}.
\end{gather}

We assume DM and DE to have a constant Equation of State (EoS) parameter, and therefore, the time component equation is the only independent equation. Taking the leading order contribution to Q from $\rho_M$ and $\rho_E$, the energy transfer is parametrized by the phenomenological model\cite{zimdahl_models_2012,2016Wang_IDE_theory}:
\begin{equation}
    Q(\rho_M,\rho_E)=\xi_1\rho_M+\xi_2\rho_E
\end{equation}

The matter evolution follows the time component of equations \ref{eq:DM_T} and \ref{eq:DE_T}:
\begin{gather}
    \dot{\rho}_M+3H\rho_M=Q(\rho_M,\rho_E),\label{eq:dm_energy_balance}\\
    \dot{\rho}_E+3H(\omega_E+1)\rho_E=-Q(\rho_M,\rho_E),\label{eq:de_energy_balance}
\end{gather}

where $\omega_E$ is the EoS parameter of DE. Following Liu et al.\cite{liu_dark_2022} we focus on a specific subset of parameter space that probes the effect of $\xi_2$, listed in table \ref{tab:IDEcosparams}.

\begin{center}
\begin{tabular}{|c|ccc|}
\hline
 Model           & IDE I  & \lcdm & IDE II\\
\hline
$\omega_E$       & -0.9191  & -1    & -1.088 \\
$\xi_2$          & -0.1107  &  0    & 0.05219\\
$\Omega_b h^2$   & 0.02223 & 0.02225 & 0.02224\\
$\Omega_c h^2$   & 0.0792   & 0.1198   & 0.1351\\
$ln (10^{10} A_s)$ & 3.099    & 3.094    & 3.097\\
$n_s$            & 0.9645   & 0.9645   & 0.9643\\
$H_0[km\ s^{-1} Mpc^{-1}]$            & 68.18    & 67.27    & 68.35\\
\hline

\end{tabular}
\captionof{table}{Model parameters considered in this work. \lcdm \ parameters come from Planck collaboration 2015 results\cite{planck_collaboration_planck_2016}). IDE I and IDE II are two set of different best fit values with Planck CMB + BAO + SNIa + $H_0$ respectively in the domain of $\omega_E<-1$ and $\omega_E>-1$\cite{costa_testing_2014,liu_dark_2022}.}
\label{tab:IDEcosparams}
\end{center}

\begin{figure}[ht]
    \centering
    \includegraphics[width=0.9\linewidth]{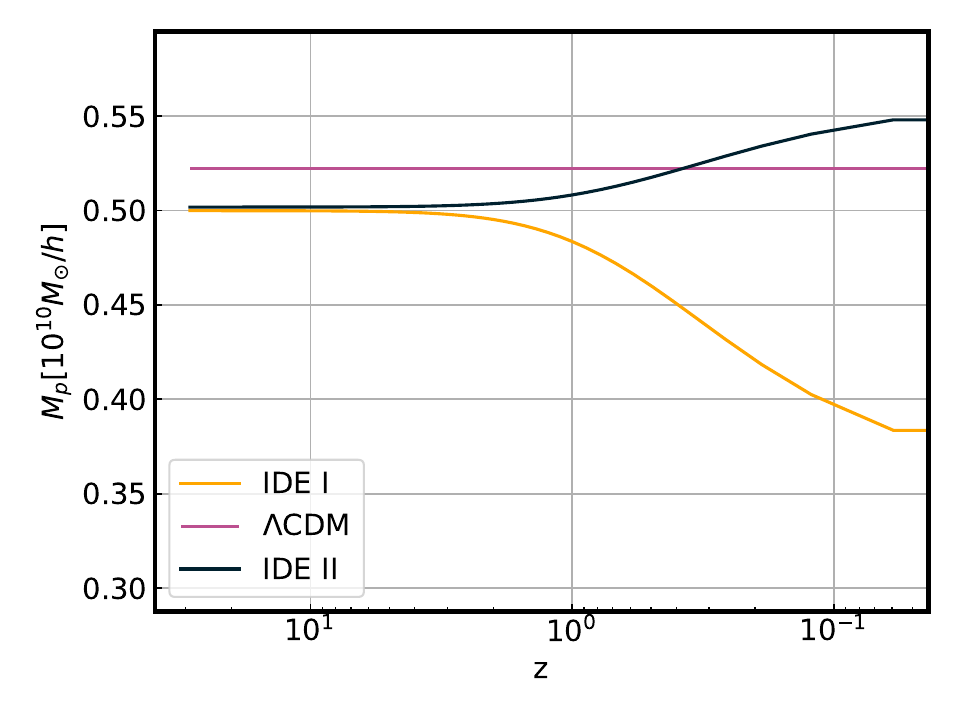}
    \caption{Mass of DM particles at different redshift in IDE I, \lcdm \ and IDE II, respectively. Note that the difference between IDE models and \lcdm \ is only significant at $z<4$, which highlights the low-redshift property of the DM-DE interaction.}
    \label{fig:dmmass}
\end{figure}

The matter equations are affected by the IDE energy transfer, and the DM energy density evolution deviates from $\propto a^{-3}$ (as in \lcdm ):
\begin{equation}
    \rho_{M}(a) = \frac{3H_0^2}{8\pi G}\left[\Omega_M+\frac{\xi_2\Omega_E}{\xi_2+\omega_E}(1-a^{-3(\omega_E+\xi_2)})\right]a^{-3},\label{eq:dm_density}
\end{equation}

and this indicates that the DM particle mass evolves during the simulation of the model IDE I and IDE II. As demonstrated in Fig. \ref{fig:dmmass}. DM particles decay in IDE I while proliferating in IDE II.

\subsection{N-body Simulation for IDE Models}\label{subsec:comp_methods}

In order to predict the large-scale structure configuration at different IDE parameters, we adopt a Dark-Matter-only N-body simulation

Our work is based on code \texttt{ME-GADGET} \cite{zhang_fully_2018} simulation which establishes a full pipeline to account for IDE effects. The \texttt{ME-GADGET} is a modified version of \texttt{GADGET-2} code \cite{GADGET_code_1,GADGET_code_2}. Our simulation box size is $200Mpc/h$ with particle number $512^3$, with periodic boundary conditions. The halos are identified with Amiga Halo Finder (AHF)\cite{AHF_halo_finder}. The pre-initial conditions are generated by the capacity-constrained Voronoi tessellation (CCVT) method\cite{2018_Liao_CCVT,2021MNRASCCVT}. The initial power spectrum is generated by a modified version of \texttt{CAMB}\cite{2002_Lewis_CMAB}. The initial condition is generated by a modified version of \texttt{2LPTic}\cite{2006_Crocce_2LPtic}.

The simulation initializes at redshift $z=49$. The initial particle mass of IDE I and IDE II are $5.0 \times 10^9 h^{-1}M_\odot$ and particle mass evolves following Fig. \ref{fig:dmmass}, the particle mass of \lcdm\ is $5.2 \times 10^9 h^{-1}M_\odot$. The comoving softening radius parameter is $4h^{-1}kpc$ in our simulation, while particle mean distance is $\sim 400 kpc$.



We identify halos with overdensity parameter $\Delta = 200$ with respect to mean background density $\bar{\rho}_M(z)$ at the corresponding redshift. The virial radii of halos are defined as $\Delta=200$ overdensity radius $R_{vir}$, and the mass of the halos is $M$. In order to suppress the particle resolution noise of halo shape and angular momentum ,a minimum requirement of $\sim 10^2$ particles is required\cite{2007MilliniumNpart}. In our work, we use halos with particle numbers larger than $N_{min}=100$. 

\begin{figure}[ht]
    \centering
    \includegraphics[width=0.9\linewidth]{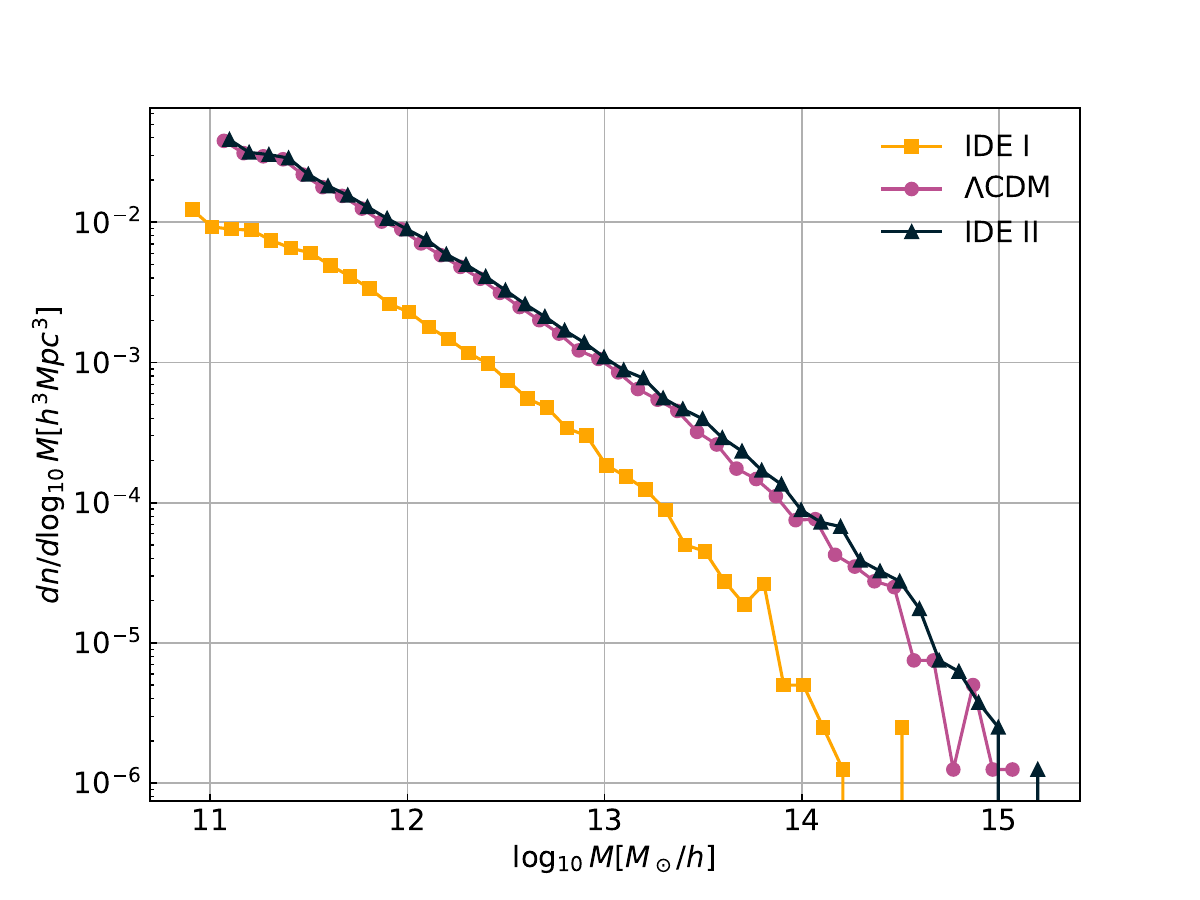}
    \caption{Halo mass functions at redshift 0 in IDE I, \lcdm\ and IDE II respectively. The total number of halos in IDE I is significantly less than \lcdm \ due to DM decay.}
    \label{fig:hmf}
\end{figure}

IDE has a significant impact on halo total count and the halo mass function, shown in Fig.\ref{fig:hmf}. DM decay in IDE I causes significantly fewer halos compared to \lcdm\ and IDE II.

In order to examine the physical relation between halo and its forming environment, the angular momentum and shape major axes of halos are computed and normalized to unit vectors.

The definition of angular momentum for each halo is:
\begin{equation}
    \vec{L}_{halo}=\sum_{i\in halo}m_{i} \vec{v'}_i\times \vec{r'}_i,
\end{equation}
where i loops through all particles inside one halo, $m_i$ is the particle mass, $\vec{r'}_i=\vec{r}_i-\vec{r}_c$ and $\vec{v'}_i=\vec{v}_i-\vec{v}_c$ are the displacement and velocity relative to the halo centre of mass.

The shape orientation is characterized by the major axes of the inertia tensor:
\begin{equation}
    \overleftrightarrow{I}_{halo} = \sum_{i \in halo} m_{i} \vec{r'}_i\otimes\vec{r'}_i,
\end{equation}
where $\vec{r'}$ are particle positions relative to the halo centre of mass. The inertia tensor is symmetric and can be diagonalized under three mutually orthogonal major axes vectors $\vec{e_a},\vec{e_b},\vec{e_c}$:
\begin{equation}
    \overleftrightarrow{I}_{halo} = \sum_{\alpha = a,b,c}I_{\alpha}\vec{e}_{\alpha}\otimes\vec{e}_{\alpha}
\end{equation}
with the eigenvalues in descending order $I_a\geq I_b\geq I_c$.

Within the scope of our study, we focus primarily on the primary major axis $\vec{G}=\vec{e}_a$ because it characterizes the most elongated direction of the halo in the gravitational environment.

\subsection{Tidal Field Eigendirections\label{Sec:tidal_intro}}

In studies of halo orientation in \lcdm, halo shape is aligned with cosmic web structures, including filaments and sheets, which is characterized by the tidal environment \cite{hahn_evolution_2007}. Therefore, it is interesting to investigate the tidal field around halos in the presence of IDE.

Studies of the halo orientation alignment in the \lcdm \ model by Wang et al.\cite{wang_build_2018} showed that halos with $z=0$ virial mass $M>10^{12} M_\odot \cdot h^{-1}$ have spin direction statistically perpendicular to the tidal field eigendirection, and $M<10^{12} M_\odot \cdot h^{-1}$ halos statistically align with tidal direction. This emphasizes the importance of the tidal field as a significant factor in halo alignment. Thus, in order to compare the effects of IDE on the halo shape and major axis, we need to introduce the tidal field eigendirections.

The tidal field is computed by definition\cite{hahn_evolution_2007,wang_spin_2018}:
\begin{equation}
    \mathrm{T}_{ij}=\partial_i \partial_j\Phi,
\end{equation}
where the gravitational potential field $\Phi$ calculated by the Poisson equation:
\begin{equation}
	\nabla^2 \Phi = 4\pi G \rho.
\end{equation}
Numerically, the mass assignment used the cloud-in-cell (CIC) method from the \texttt{Pylians}\cite{Pylians} package. The assigned 3d mass grid has a spacing of $0.781Mpc/h$. A Gaussian smoothing was carried out in Fourier space with a smoothing radius of $R_s=2Mpc/h$. We use this smoothing radius consistent with Wang et al.\cite{wang_general_2017}.

The smoothed density field was then convolved with the 3D Green's function to obtain the gravitational potential field:
\begin{equation}
	\Phi(\vec{r}) = -\int d^3r' \frac{G\rho(\vec{r'})}{|\vec{r}-\vec{r'}|}.
\end{equation}

The tidal shear tensor is the traceless part of the Hessian matrix of the gravitational potential field:
\begin{equation}
    \mathcal{T}_{ij}(\vec{r})=\partial_i\partial_j \Phi(\vec{r})-\frac{1}{3}\delta_{ij}\sum_{k}\partial_k^2\Phi(\vec{r}) 
\end{equation}

The tidal field direction $\vec{T}_3$ ($\vec{T}$ in the following) is the eigenvector of $\mathcal{T}_{ij}$ with the least eigenvalue. This direction also corresponds to the direction where dark matter collapses the least in linear perturbation theories.

\begin{figure*}[ht]
    \centering
    \includegraphics[width=1.0\linewidth]{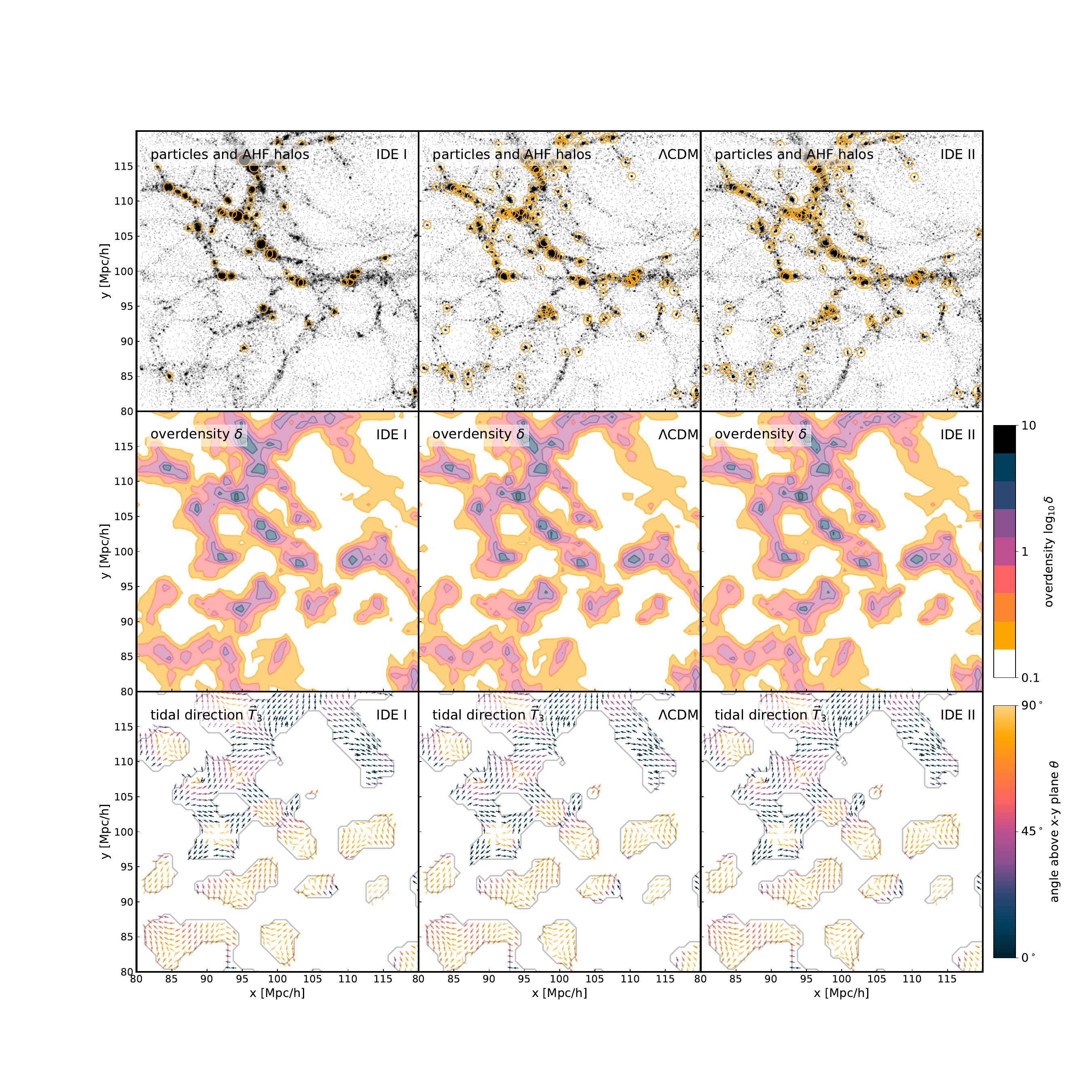}
    \caption{The figure shows simulated particle snapshots (first row), density heightmap (second row), and the tidal field direction $\vec{T}$ (third row) in models IDE I, \lcdm \ and IDE II. First row: the simulation particle snapshots at redshift 0 (black point cloud), and the AHF halos (orange circles with radius representing the viral radius $R_{vir}$). Second row: density heatmap at overdensity $\delta>0$. Third row: the tidal tensor least eigenvalue arrow plot in dense regions, with coloring denoting the inclination angle above the x-y plane. }
    \label{fig:tidal_demo}
\end{figure*}

From our simulation result, a slice comparison of the particles, halos, matter density, and tidal vectors is shown in Fig. \ref{fig:tidal_demo}. Row 1-3 corresponds to particle+halos, dark matter overdensity, and tidal field direction $\vec{T}$, respectively.


In Fig.\ref{fig:tidal_demo}, comparing the second row to the third row, we can see that the tidal field follows the direction of filaments in filamentary structures. Inside filaments, $\vec{T}$ points towards the direction of the filament within the perspective plane (colored in dark blue), and at filament nodes, $\vec{T}$ points above the perspective plane (colored in yellow).

\section{Result\label{result}}
\begin{figure*}
\includegraphics[width=0.95\textwidth]{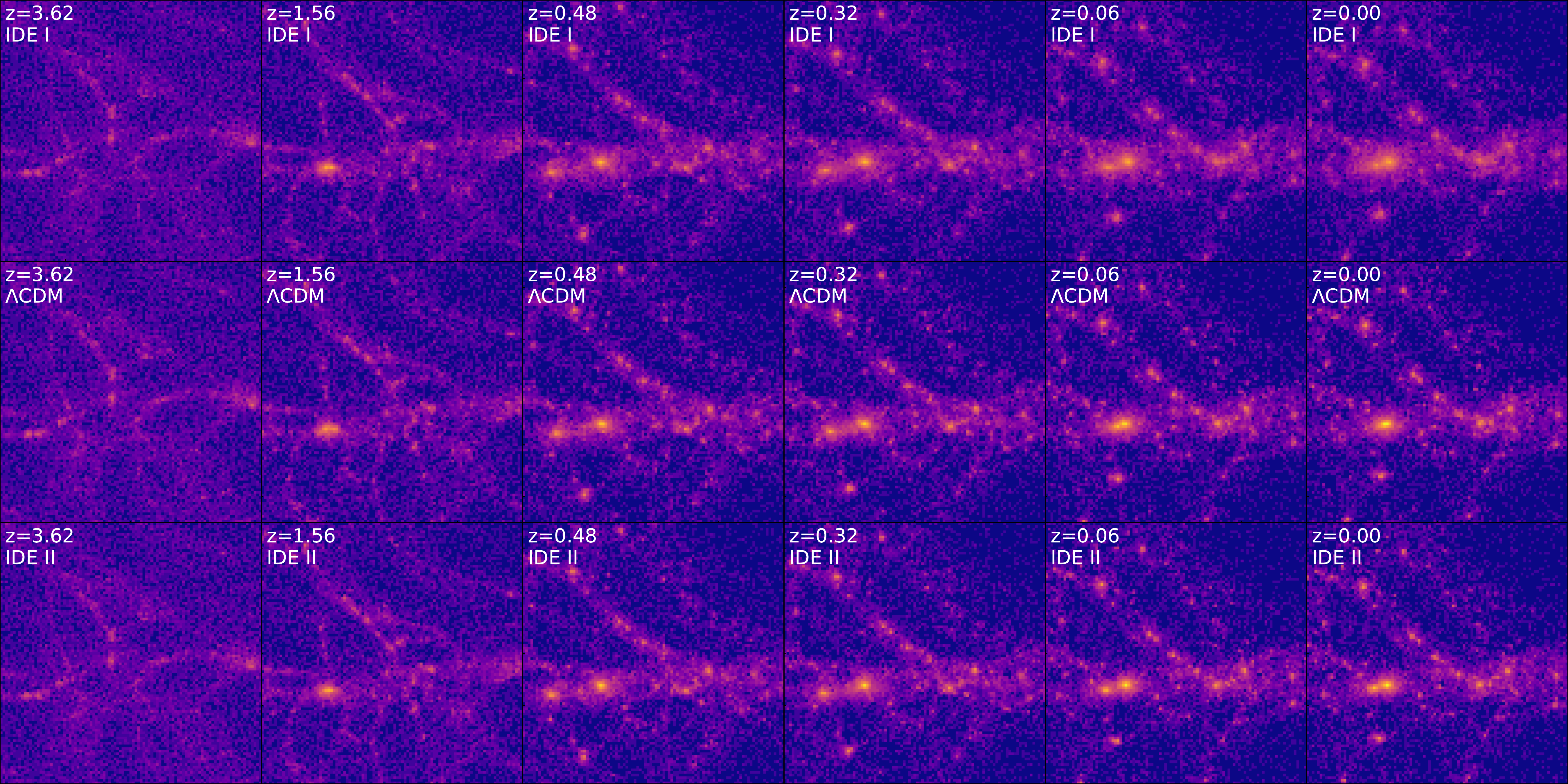}
\caption{This figure illustrates the evolution of a single halo in its large-scale environment. From left to right are the snapshots from $z=3.62$ to $z=0.0$. The width and the height of the snapshots are both $10Mpc/h$, and the depth of the snapshots is $10Mpc/h$. Compared to $\Lambda CDM$, the halos in IDE I are more vulnerable against merging and the tidal environment due to its loose structure, while the halos in IDE II are more robust.}
\label{fig:animation}
\end{figure*}
We have seen that IDE has a significant impact on the halo mass function and halo concentration. The underlying physics is that the formation history of halos is changed due to the interaction between dark matter and dark energy. We have shown the evolution history of a halo in its large-scale environment in Fig.~\ref{fig:animation}. We can see that, compared to $\Lambda CDM$, the halos in IDE I are more vulnerable against merging and the tidal environment due to its loose structure, while the halos in IDE II are more robust, which is expected. We will quantify such effects and provide fitting formulas, so that we can use them in further modelling. 

Building on the framework outlined in the previous section, we now present the quantitative results for halo-tidal field alignments. In order to assess how IDE models modify the strength and scale of alignment compared to \lcdm, we examine the in-situ spin-tidal in section \ref{subsec:in-situ-spin} and shape-tidal correlations in section \ref{subsec:in-situ-shape}. Besides, we also examine the alignment correlation as a function of halo separation, and present our results in section \ref{subsec:corrfunc}. This focus enables us to isolate the signatures of IDE on halo orientation and angular momentum.

The alignment strength between two 3D vectors is defined as follows in this work:
\begin{equation}
	\eta_{uv} = (\vec{u}\cdot\vec{v})^2-1/3.\label{eq:alignment-strength}
\end{equation}
The notation $u,v$ stands for halo quantities: 
\begin{enumerate}
    \item Halo shape $\vec{G}$, as discussed in \ref{subsec:comp_methods}. The vector is normalized to unit length.
    \item Halo angular momentum (spin) $\vec{L}$, as discussed in \ref{subsec:comp_methods}. The vector is normalized to unit length.
    \item Tidal field experienced by halo $\vec{T}$, as discussed in \ref{Sec:tidal_intro}. The tidal field for the individual halo is calculated by the $\vec{T}$ tidal eigenvector direction at the closest gridpoint.
\end{enumerate}

The $-1/3$ constant in Equation \ref{eq:alignment-strength} ensures that the expectation value for the alignment strength is 0 for a spherically random sample. For halo samples, an averaged positive alignment strength stands for a tendency to align, and a negative alignment strength indicates a tendency to be perpendicular. This alignment definition is invariant under the inversion of either one of the vectors and is thus robust under the $\pm$ ambiguity of eigenvector directions.

\subsection{Spin-tidal In-situ Correlation\label{subsec:in-situ-spin}}

In \lcdm, the halo spin-tidal alignment depends on halo mass, with the alignment switching from positive to negative at  $M\sim 10^{12}M_\odot \cdot h^{-1}$ \cite{wang_build_2018}. Here we examine the behavior of IDE models and compare them to previous studies of \lcdm. We examine halos within mass range $10^{11}M_\odot\cdot h^{-1} \sim 10^{13}M_\odot\cdot h^{-1}$ in IDE I, IDE II and \lcdm. We take the average of $\eta_{LT}=(\vec{L}\cdot\vec{T})^2-1/3$ of equal-sample mass bins in each of the models\footnote{The IDE I model has fewer mass bins due to its fewer total halo counts.} and show the relation of $\langle\eta_{LT}\rangle (M)$.

\begin{figure*}[ht]
    \centering
    \includegraphics[width=1.0\linewidth]{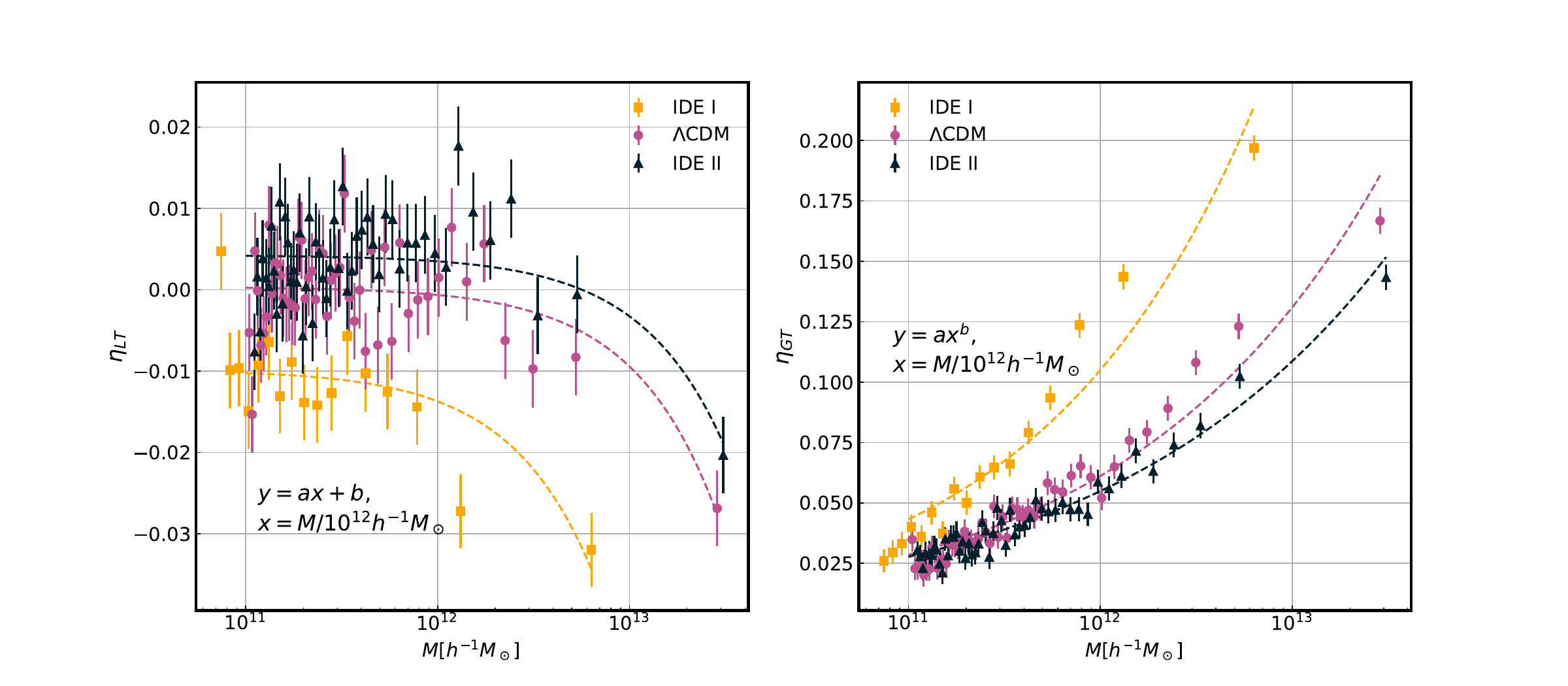}
    \caption{The figure shows halo alignment strength vs halo virial mass. Left panel: halo spin-tidal alignment, linear fit lines are plotted as dashed lines. Right panel: halo shape-tidal alignment, powerlaw fit curves are plotted as the dashed lines. The best-fit parameters and fit forms are listed in Table \ref{tab:insitu_corr_fits}.}
    \label{fig:insitu_correlation}
\end{figure*}

The spin-tidal relation with halo mass is shown in the left panel of Fig. \ref{fig:insitu_correlation}. In all three models, alignment strength generally decreases with halo mass, and the zero crossing happens at $\sim 10^{12}h^{-1}M_\odot$ in \lcdm, consistent with the results from Wang et al.\cite{wang_build_2018}. This means smaller (massive) halos tend to have spin parallel (perpendicular) to the tidal field, respectively. The linear best-fit parameters are listed in Table \ref{tab:insitu_corr_fits}. The fitting parameter shows a significant tendency for the spin to be perpendicular (parallel) to the tidal field for IDE I (IDE II), compared to \lcdm \ model. This implies DM mass decay in IDE I makes the halo spin tend to be perpendicular to the tidal field eigendirection.




\subsection{Shape-tidal In-situ Correlation\label{subsec:in-situ-shape}}

In \lcdm, the halo shape tends to align with the tidal field, which is caused by both tidal forces and feeding mechanisms, resulting in a shape elongated towards tidal eigendirection. To see the effect of IDE on shape-tidal alignment, we use the same binning scheme as \ref{subsec:in-situ-spin} and examine the average halo alignment strengths $\langle \eta_{GT}\rangle(M)$ in mass range $10^{11}M_\odot\cdot h^{-1} \sim 10^{13}M_\odot\cdot h^{-1}$.

We show the shape-tidal alignment dependence on halo mass in the right panel of Fig.\ref{fig:insitu_correlation}. Alignment between shape and tidal field is positive across three models. This indicates that in IDE models, the halo shapes are aligned with tidal direction $\vec{T}$ as in \lcdm. A power law fit is given with $ y = ax^b $, where $ x = M / 10^{12} M_\odot \cdot h^{-1}$, and the corresponding fit parameters are listed in Table \ref{tab:insitu_corr_fits}. Comparison between models shows that generically, the shapes of IDE I (IDE II) halos are more (less) parallel to the tidal field than \lcdm.




\begin{table*}[t]
\centering
\small
\caption{
Best-fit parameters for the in-situ alignment correlations between halo spin and tidal field ($\eta_{LT}$) 
and between halo shape and tidal field ($\eta_{GT}$) across cosmological models.
The fitting forms are:
$\eta_{LT}(M) = a\,x + b$, where $x = M / 10^{12} M_\odot\cdot h^{-1}$;
and $\eta_{GT}(M) = a\,x^b$, where $x = M / 10^{12} M_\odot\cdot h^{-1}$.
Uncertainties correspond to $1\sigma$ posterior standard deviations from MCMC unless otherwise noted.
}
\vspace{4pt}
\begin{tabular}{lcccc}
\hline
Corr. & Model & Form & $a$ & $b$  \\
\hline
Spin–Tidal & IDE I   & $a\,x + b$ & $(-3.88 \pm 0.77)\times 10^{-3}$ & $(-9.83 \pm 1.25)\times 10^{-3}$  \\
Spin–Tidal & \lcdm    & $a\,x + b$ & $(-0.98 \pm 0.16)\times 10^{-3}$ & $(0.38 \pm 0.71)\times 10^{-3}$   \\
Spin–Tidal & IDE II  & $a\,x + b$ & $(-0.75 \pm 0.15)\times 10^{-3}$ & $(4.22 \pm 0.69)\times 10^{-3}$   \\
\hline
Shape–Tidal & IDE I   & $a\,x^b$ & $0.1050±0.0016$ & $0.386±0.009$  \\
Shape–Tidal & \lcdm    & $a\,x^b$ & $0.0612±0.0008$ & $0.331±0.007$  \\
Shape–Tidal & IDE II  & $a\,x^b$ & $0.0549±0.0008$ & $0.2969±0.008$  \\
\hline
\end{tabular}
\label{tab:insitu_corr_fits}
\end{table*}

\subsection{Alignment Correlation\label{subsec:corrfunc}}

In-situ alignment probes the direct coupling between halo quantities and the surrounding tidal environment. Now we extend our analysis across spatial separations using alignment correlation functions. The alignment correlation function quantifies how the alignment strength persists or decorrelates with spatial separation between halos. This approach allows us to capture the large-scale coherence of halo-tidal interactions.

The quantitative alignment correlation function is defined as the binned average of alignment strength $\eta_{uv}$ between halo pairs as a function of halo separation distance $r$.

The correlation is calculated from the following steps:
\begin{enumerate}
    \item Setup equal-spacing radius bins in range $0.1Mpc\cdot h^{-1}\sim 20Mpc\cdot h^{-1}$.
    \item Find all halo pairs from the \texttt{AHF} halo catalog (see section \ref{subsec:comp_methods}) that have separation in each of the radius bins.
    \begin{equation*}
        pairs(r) = \left\{[halo1,halo2]|r_{min}^{bin}<d_{halo1,halo2}<r_{max}^{bin}\right\}
    \end{equation*}
    \item Calculate the alignment strength of halo quantities $\vec{u},\vec{v}$ (one of shape, spin, tidal vectors $G,L,T$), defined as:
    \begin{equation*}
        \eta_{u,v}^{pair}=(\vec{u}_{halo1}\cdot \vec{v}_{halo2})^2-1/3.
    \end{equation*}
    \item Perform statistics within each bin, and obtain the mean and standard error of alignment strength.
    \begin{align*}
        \eta_{u,v}(r)&=\frac{\sum_{pair\in pairs(r)}\eta_{u,v}^{pair}}{N_{pairs}(r)}.\\
        \sigma_{uv}^2(r)&=\frac{1}{N_{pairs}(r)-1}\left[\frac{\sum_{pair\in pairs(r)}(\eta_{u,v}^{pair})^2}{N_{pairs}(r)}-\eta_{u,v}(r)^2\right]
    \end{align*}
    where $N_{pairs}(r)$ is the pair count within each bin.
\end{enumerate}



\begin{figure*}[t]
  \centering
  \includegraphics[width=\textwidth]{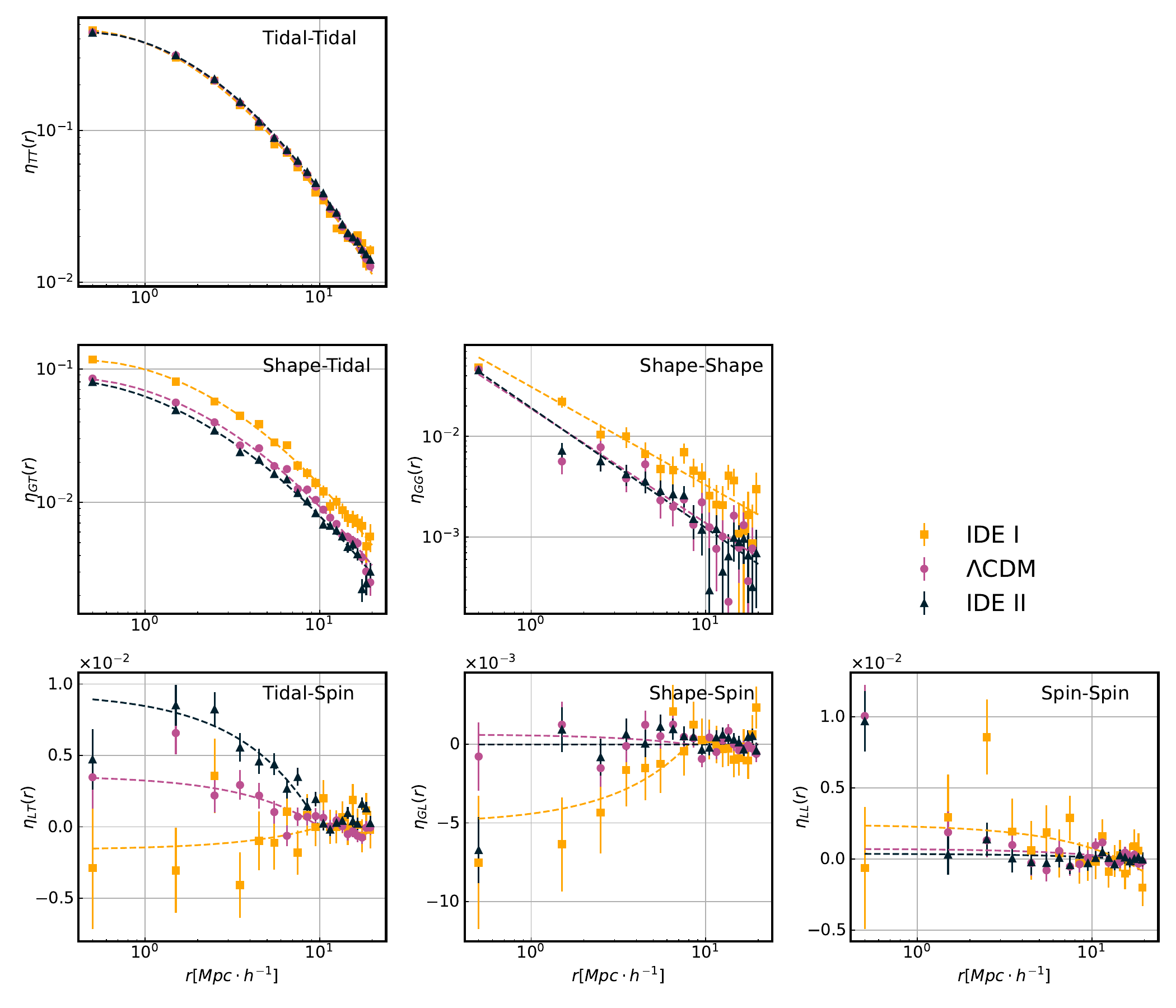}
  \caption{Comparison of alignment correlation functions versus comoving separation \(r\) (in \(h^{-1}\,Mpc\)). Top: tidal–tidal (\(\eta_{TT}\)). Middle: shape–tidal (\(\eta_{GT}\)), and shape–shape (\(\eta_{GG}\)). Bottom: tidal–spin (\(\eta_{LT}\)), shape–spin (\(\eta_{GL}\)), and spin–spin (\(\eta_{LL}\)). Panel-specific scale factors (e.g., \(\times 10^{-2}\), \(\times 10^{-3}\)) indicate the plotting range. The fitting form and parameters are listed in Table \ref{tab:all_corr_fits}.}
  \label{fig:corrfunc_comparison}
\end{figure*}

Fig. \ref{fig:corrfunc_comparison} shows the IDE model comparison of TT, GT, LT, GG, GL, LL cross and auto correlations. The model comparison shows the following qualitative results:
\begin{enumerate}
    \item Tidal autocorrelation $\eta_{TT}(r)$ shows consistency across IDE models. This implies that the positions of Large-Scale Structures like filaments and clusters are less affected by IDE models.
    \item Under consistent tidal environments, the shape-tidal crosscorrelation $\eta_{GT}(r)$ is stronger in IDE I model and weaker in IDE II model compared to \lcdm.
    \item Tidal-spin crosscorrelation $\eta_{LT}(r)$ is stronger for IDE II and weaker for IDE I than \lcdm.
    \item Shape autocorrelation $\eta_{GG}(r)$ is stronger for IDE I, than IDE II and \lcdm. The power-law fit radius power is also higher for IDE. This implies that the shape correlation persists for longer distances for IDE.
    \item Shape-spin crosscorrelation $\eta_{GL}(r)$ and spin autocorrelation $\eta_{LL}(r)$ have insignificant statistical difference across IDE models.
\end{enumerate}



The strengthened shape-shape, shape-tidal crosscorrelation in IDE I compared to IDE II and \lcdm is consistent with the in-situ shape-tidal crosscorrelation discussed in section \ref{subsec:in-situ-shape}. Combined with a consistent tidal alignment correlation across models. This implies that a stronger coupling of halos to the tidal environment in IDE I model.

The strengthened spin-tidal crosscorrelation in IDE II model is consistent with the in-situ spin-tidal crosscorrelation discussed in section \ref{subsec:in-situ-spin}.

\begin{table*}[t]
\centering
\small
\caption{
Best-fit parameters for all alignment–correlation functions across cosmological models. 
The fitting forms are:
TT, GT: $\eta_{TT,GT}(r)=\exp\!\big(a+b\ln r+c(\ln r)^2\big)$;
LT: $\eta_{LT}(r)=a\,(r-10)\,\Theta(10-r)$;
GG: $\eta_{GG}(r)=a\,r^{\,b}$;
GL: $\eta_{GL}(r)=a\,(r-8)\,\Theta(8-r)$;
LL: $\eta_{LL}(r)=a\,r+b$.
Distances $r$ are in $h^{-1}\,\mathrm{Mpc}$, parameters are dimensionless, and $\Theta$ denotes the Heaviside step function.
Uncertainties correspond to the $1\sigma$ posterior standard deviations from MCMC unless otherwise noted.
}
\vspace{4pt}
\begin{tabular}{lccccc}
\hline
Corr. & Model & Form & $a$ & $b$ & $c$ \\
\hline
TT    & IDE I   & $\exp(a+b\ln r+c(\ln r)^2)$ & $-0.9746 \pm 0.0065$ & $-0.4469 \pm 0.0081$ & $-0.2419 \pm 0.0046$ \\
TT    & \lcdm    & $\exp(a+b\ln r+c(\ln r)^2)$ & $-0.9720 \pm 0.0031$ & $-0.4084 \pm 0.0040$ & $-0.2503 \pm 0.0020$ \\
TT    & IDE II  & $\exp(a+b\ln r+c(\ln r)^2)$ & $-0.9696 \pm 0.0031$ & $-0.3984 \pm 0.0039$ & $-0.2484 \pm 0.0019$ \\
\hline
GT    & IDE I   & $\exp(a+b\ln r+c(\ln r)^2)$ & $-2.312 \pm 0.024$ & $-0.372 \pm 0.032$ & $-0.213 \pm 0.015$ \\
GT    & \lcdm    & $\exp(a+b\ln r+c(\ln r)^2)$ & $-2.675 \pm 0.016$ & $-0.417 \pm 0.021$ & $-0.197 \pm 0.009$ \\
GT    & IDE II  & $\exp(a+b\ln r+c(\ln r)^2)$ & $-2.777 \pm 0.018$ & $-0.474 \pm 0.023$ & $-0.183 \pm 0.010$ \\
\hline
LT    & IDE I   & $a\,(r-10)\,\Theta(10-r)$ & $(+1.6\pm 1.3)\times10^{-4}$ & -- & -- \\
LT    & \lcdm    & $a\,(r-10)\,\Theta(10-r)$ & $(-3.6\pm 0.6)\times10^{-4}$ & -- & -- \\
LT    & IDE II  & $a\,(r-10)\,\Theta(10-r)$ & $(-9.4\pm 0.6)\times10^{-4}$ & -- & -- \\
\hline
GG    & IDE I   & $a\,r^{\,b}$ & $0.0310 \pm 0.0024$ & $-0.974 \pm 0.059$ & -- \\
GG    & \lcdm    & $a\,r^{\,b}$ & $0.0188 \pm 0.0012$ & $-1.134 \pm 0.053$ & -- \\
GG    & IDE II  & $a\,r^{\,b}$ & $0.0192 \pm 0.0012$ & $-1.192 \pm 0.056$ & -- \\
\hline
LG    & IDE I   & $a\,(r-8)\,\Theta(8-r)$  & $(+6.3\pm 2.1) \times10^{-4}$ & -- & -- \\
LG    & \lcdm    & $a\,(r-8)\,\Theta(8-r)$  & $(-8.0\pm 9.8)\times10^{-5}$ & -- & -- \\
LG    & IDE II  & $a\,(r-8)\,\Theta(8-r)$  & $(0.0\pm 9.5)\times 10^{-5}$                 & -- & -- \\
\hline
LL    & IDE I   & $a\,r+b$ & $(-1.68\pm 0.68)\times10^{-4}$ & $(2.43\pm0.91)\times10^{-3}$ & -- \\
LL    & \lcdm    & $a\,r+b$ & $(-4.5\pm2.9)\times10^{-5}$  & $(7.2\pm3.9)\times10^{-4}$  & -- \\
LL    & IDE II  & $a\,r+b$ & $(-2.5\pm2.7)\times10^{-5}$  & $(3.7\pm3.8)\times10^{-4}$  & -- \\
\hline
\end{tabular}
\label{tab:all_corr_fits}
\end{table*}

\section{Conclusion and discussion\label{discussion}}

This study compares dissipative and proliferating IDE models (IDE I and IDE II) in terms of halo properties, specifically the correlation between halo spin, shape, and the surrounding tidal field.

We find that there are fewer halos with spin aligned to the tidal field and more halos with shape aligned to the tidal field in the dissipative IDE model. In contrast, proliferating IDE models show the opposite trend, with more halos having spin aligned to the tidal field and fewer halos with shape aligned. This difference in shape-tidal alignment results in the shape-shape autocorrelation significantly enhanced in dissipative IDE, while it is mildly reduced in proliferating IDE models.

We highlight the following conclusions we find in the study:
\begin{itemize}
    \item Spin-tidal alignment: stronger in IDE II, weaker in IDE I, as demonstrated in Fig. \ref{fig:insitu_correlation} left panel.
    \item Shape-tidal alignment: stronger in IDE I, weaker in IDE II, as demonstrated in Fig. \ref{fig:insitu_correlation} right panel.
    \item Tidal-tidal alignment: tidal directional autocorrelation is consistent across IDE models, which implies that the IDE models have less effect on the location and direction of Large-Scale Structures. The slice comparison is shown in Fig. \ref{fig:tidal_demo}, and the autocorrelation is shown in upper-left panel of Fig. \ref{fig:corrfunc_comparison}.
    \item Shape-shape alignment: significantly stronger in IDE I and weaker in IDE II, as demonstrated in the central panel of Fig. \ref{fig:corrfunc_comparison}.
    \item Spin-shape, spin-spin alignment: the spin-shape and spin-spin correlations are statistically insignificant, as shown in the bottom row of Fig. \ref{fig:corrfunc_comparison}. This spin alignment signal is weak relative to the data scatter and could be further investigated in future simulations with larger box sizes and higher particle counts.
\end{itemize}

The effect of higher (lower) shape-tidal and shape-shape correlation in dissipative (proliferating) IDE models can be explained by the difference in the density profile. Liu et al.\cite{liu_dark_2022} found that the halo concentrations $c$ in dissipative IDE models are lower and halos are more eccentric than in proliferating IDE and \lcdm. This makes halos in dissipative IDE more sensitive to environmental effects from the tidal field, leading to stronger shape-tidal and shape-shape correlations.

To account for the effect of lower spin-tidal correlation seen in Fig. \ref{fig:insitu_correlation} left panel and Fig. \ref{fig:corrfunc_comparison} lower left panel, we postulate that, in dissipative IDE, the halos at z=0 could only have survived the mass decay and halo disassembly by accreting material from dense regions (clusters and filaments). As Wang et al. discussed, the halo migration while staying within the filament helps gain angular momentum perpendicular to tidal eigendirection \cite{wang_build_2018}, which means smaller mass halos in dissipative IDE correspond to more massive halos in \lcdm \ that stay extended periods inside filaments.

Our findings provide the ingredients for a more comprehensive modeling for the IA calibration of weak lensing measurements for the CSST, including the IDE effects on inter-halo and halo-LSS alignment. The quantitative characterization of IDE effects on IA is beyond the scope of this work, and we plan to do it in future work. The future work includes more simulations to cover the parameter space and construct an emulator or a fitting model for building the IA calibration model, a careful choice of the related parameters, and an easy-to-use package will be developed.






\section{Acknowledgement}
This work was supported by National Natural Science Foundation of China (NSFC) grants 12473003 and the China Manned Space Project with no. CMS-CSST-2021-A03. JY acknowledges the support from National Natural Science Foundation of China (NSFC) grants 12203084. PW acknowledge the support from National Natural Science Foundation of China (NSFC) grants 12473009 and Shanghai Rising-Star Program No. 24QA2711100. This work was inspired by the discussion in HOUYI workshop held in Shanghai, 2023. This work is supported by the China Manned Space Program with grant no. CMS-CSST-2025-A03.

\appendix
\pagebreak


\bibliography{main}
\bibliographystyle{unsrt}
\end{document}